\documentclass[apj]{emulateapj}
\def\gtorder{\mathrel{\raise.3ex\hbox{$>$}\mkern-14mu
             \lower0.6ex\hbox{$\sim$}}}
\def\ltorder{\mathrel{\raise.3ex\hbox{$<$}\mkern-14mu
             \lower0.6ex\hbox{$\sim$}}}

\slugcomment{Accepted 2010 September 27}

\shorttitle{Supernova PTF\,09uj}

\shortauthors{Ofek et al.}

\begin{document}

\title{Supernova PTF\,09uj: A possible shock breakout from a dense circumstellar wind}
\author{E.~O.~Ofek\altaffilmark{1}$^{,}$\altaffilmark{2},
I.~Rabinak\altaffilmark{3},
J.~D.~Neill\altaffilmark{1},
I.~Arcavi\altaffilmark{3},
S.~B.~Cenko\altaffilmark{4},
E.~Waxman\altaffilmark{3},
S.~R.~Kulkarni\altaffilmark{1},
A.~Gal-Yam\altaffilmark{3},
P.~E.~Nugent\altaffilmark{5},
L.~Bildsten\altaffilmark{6}$^{,}$\altaffilmark{7},
J.~S.~Bloom\altaffilmark{4},
A.~V.~Filippenko\altaffilmark{4},
K.~Forster\altaffilmark{1},
D.~A.~Howell\altaffilmark{9}$^{,}$\altaffilmark{6},
J.~Jacobsen\altaffilmark{5}, 
M.~M.~Kasliwal\altaffilmark{1},
N.~Law\altaffilmark{10}$^{,}$\altaffilmark{1},
C.~Martin\altaffilmark{1},
D.~Poznanski\altaffilmark{4}$^{,}$\altaffilmark{5}$^{,}$\altaffilmark{2},
R.~M.~Quimby\altaffilmark{1},
K.~J.~Shen\altaffilmark{6},
M.~Sullivan\altaffilmark{11},
R.~Dekany\altaffilmark{12},
G.~Rahmer\altaffilmark{12},
D.~Hale\altaffilmark{12},
R.~Smith\altaffilmark{12},
J.~Zolkower\altaffilmark{12},
V.~Velur\altaffilmark{12},
R.~Walters\altaffilmark{12},
J.~Henning\altaffilmark{12},
K.~Bui\altaffilmark{12}, and
D.~McKenna\altaffilmark{12}
}

\altaffiltext{1}{Division of Physics, Mathematics and Astronomy,
  California Institute of Technology, Pasadena, CA 91125.}
\altaffiltext{2}{Einstein Fellow.}
\altaffiltext{3}{Benoziyo Center for Astrophysics, Weizmann Institute
  of Science, 76100 Rehovot, Israel.}
\altaffiltext{4}{Department of Astronomy, University of California,
  Berkeley, Berkeley, CA 94720-3411.}
\altaffiltext{5}{Lawrence Berkeley National Laboratory, 1 Cyclotron
  Road, Berkeley, CA 94720.}
\altaffiltext{6}{Department of Physics, Broida Hall, University of
  California, Santa Barbara, CA 93106.}
\altaffiltext{7}{Kavli Institute for Theoretical Physics, Kohn Hall,
  University of California, Santa Barbara, CA 93106.}
\altaffiltext{8}{Spitzer Science Center, MS 220-6, California
  Institute of Technology, Jet Propulsion Laboratory, Pasadena, CA
  91125.}
\altaffiltext{9}{Las Cumbres Observatory Global Telescope Network,
  6740 Cortona Dr., Suite 102, Goleta, CA 93117.}
\altaffiltext{10}{Dunlap Institute for Astronomy and Astrophysics,
  University of Toronto, 50 St. George Street, Toronto, Ontario M5S
  3H4, Canada.}
\altaffiltext{11}{Department of Physics, University of Oxford, Denys
  Wilkinson Building, Keble Road, Oxford OX1 3RH, UK.}
\altaffiltext{12}{Caltech Optical Observatories, California Institute
  of Technology, Pasadena, CA 91125.}

\begin{abstract}

%        1         2         3         4         5         6         7         8   
%23456789 123456789 123456789 123456789 123456789 123456789 123456789 1234567890

Type-IIn supernovae (SNe), which are characterized by strong interaction of their
ejecta with the surrounding circumstellar matter (CSM),
provide a unique opportunity to study the mass-loss history of
massive stars shortly before their explosive death.
We present the discovery and follow-up observations of
a Type IIn SN, PTF\,09uj,
detected by the Palomar Transient Factory (PTF).
Serendipitous observations by {\it GALEX} at
ultraviolet (UV) wavelengths detected the rise of the SN
light curve prior to the PTF discovery.
The UV light curve of the SN rose fast,
with a time scale of a few days,
to a UV absolute AB magnitude of about $-19.5$.
Modeling our observations, we suggest that 
the fast rise of the UV light curve is due to
the breakout of the SN shock through the dense CSM
($n \approx 10^{10}$\,cm$^{-3}$).
Furthermore, we find that
prior to the explosion the progenitor went through a phase of
high mass-loss rate ($\sim 0.1$\,M$_{\odot}$\,yr$^{-1}$)
that lasted for a few years.
The decay rate of this SN was fast relative to that of other SNe~IIn.

\end{abstract}


\section{Introduction}
\label{sec:Introduction}

Early ultraviolet (UV) detection of supernovae (SNe) of all types
in the shock-breakout phase holds great potential
for probing the nature and properties of SN progenitors
(e.g., Colgate 1974; Falk \& Arnett 1977; Klein \& Chevalier 1978;
Matzner \& McKee 1999;
Waxman et al. 2007;
Katz, Budnik \& Waxman 2010;
Piro, Chang \& Weinberg 2010;
Rabinak \& Waxman 2010;
Nakar \& Sari 2010).
Shock-breakout observations
provide a measure of the progenitor radius
and the structure of its outer layers
(e.g., Soderberg et al. 2008; Gezari et al. 2008; Schawinski et al. 2008).

Type IIn SNe are characterized by the presence of a blue continuum and
narrow emission lines in their optical spectra 
(e.g., Schlegel 1990; Filippenko 1997).
These features are usually interpreted as the signatures of interaction
of the SN ejecta with dense circumstellar matter (CSM),
due to stellar mass loss prior to the explosion.
SNe~IIn are probably an
inhomogeneous class of objects whose main properties are
dictated by the presence of dense CSM
rather than by the details of the explosion.
Therefore, SNe showing evidence for strong interaction
are unique probes of scenarios in which
a stellar explosion follows a major mass-loss event,
and they can be used to study the mass-loss rate from progenitor stars
(e.g., Chugai \& Danziger 1994;
Chugai et al. 1995;
Fransson et al. 2002;
Gal-Yam et al. 2007;
Ofek et al. 2007;
Smith et al. 2007;
Smith \& McCray 2007;
Gal-Yam \& Leonard 2009;
Dessart et al. 2009).

Here we present the discovery of a Type IIn SN,
PTF\,09uj, which was serendipitously 
observed in the UV by {\it GALEX} (Martin et al. 2005) shortly after
the explosion.

\section{Observations and data Reduction}
\label{sec:Observations}

\begin{figure}
\centerline{\includegraphics[width=8.5cm]{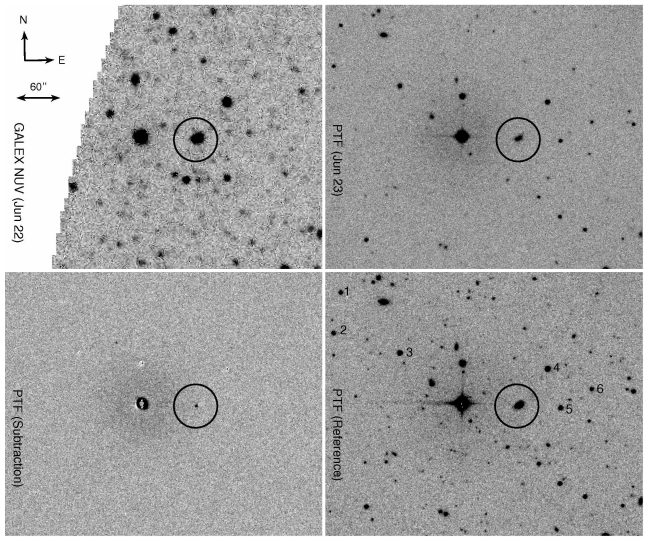}}
\caption{Images of the field of PTF\,09uj.
The SN was detected by {\it GALEX} in the $NUV$ band on June 22 (upper left),
and by PTF in the $R$ band on June 23 (upper right). The bottom-right
panel shows a reference $R$-band image prepared from PTF images
obtained before the explosion. The numbered stars (1--6) mark
the SDSS photometric reference stars. The bottom-left panel
shows the discovery of PTF\,09uj using image subtraction of PTF data.
A $30''$-radius circle marks the position of the SN in all panels.
The position of the SN is $\alpha$(J2000) = 14$^{h}$20$^{m}$11.15$^{s}$,
$\delta$(J2000) = +53$^{\circ}$33$'$41.0$''$, which is $2.7''$ from
the center of SDSS\,J142010.86+533341.9 at a position angle of 
$\approx 110$\,deg.
\label{fig:PTF09uj_Images}}
\end{figure}

SN PTF\,09uj was discovered
on 2009 June 23.30 (UTC dates are used throughout this paper) by
the Palomar Transient Factory\footnote{http://www.astro.caltech.edu/ptf/.}
(PTF; Law et al. 2009; Rau et al. 2009) conducted with the Oschin
48-inch Schmidt telescope (P48) at Palomar Observatory.
The SN is associated with an $r=17.3$\,mag
galaxy, SDSS\footnote{Sloan Digital Sky Survey; 
York et al. (2000).}\,J142010.86+533341.9 (Fig.~\ref{fig:PTF09uj_Images}) ---
a late-type disk galaxy at a redshift of $z = 0.0650 \pm 0.0001$
(distance $d \cong 292$\,Mpc\footnote{We assume WMAP-5 cosmological
 parameters (Komatsu et al. 2008) and insignificant peculiar velocity.}).

\subsection{Photometry}

Follow-up photometry was obtained using the 
automated Palomar 60-inch telescope (Cenko et al. 2006; Table~\ref{tab:Obs}).
As noted above, this field was observed by {\it GALEX} 
on several occasions (Table~\ref{tab:Obs}).

Photometry of PTF\,09uj in the PTF and P60 images
was obtained by the common point-spread function
image-subtraction method (Gal-Yam et al. 2004; Gal-Yam et al. 2008).
Estimates of uncertainties were obtained
from the scatter in the magnitudes of artificial sources.
The photometry was calibrated using $r$-band magnitudes of
six SDSS stars (Fig.~\ref{fig:PTF09uj_Images}).
%We note that the SDSS $r$-band and P48 $R$-band transmission
%curves are similar (Law et al. 2009).
Calibration errors
were added in quadrature to the image-subtraction
errors (Table~\ref{tab:Obs}).

\begin{figure}
\centerline{\includegraphics[width=8.5cm]{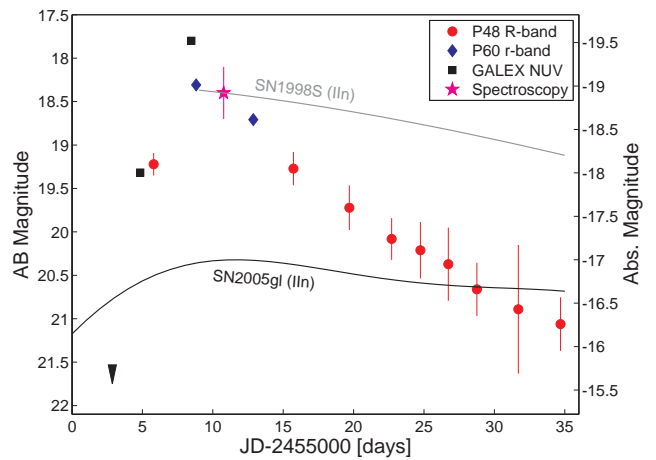}}
\caption{The light curve of PTF\,09uj from the P48 
({\it red circles}), P60 ({\it blue diamonds}),
synthetic photometry ({\it magenta star}),
and {\em GALEX} {\it NUV} observations ({\it black squares}).
The tip of the {\it black triangle}
marks the magnitude lower limit from {\em GALEX}.
The curves show the
scaled and smoothed light curves of two other SNe~IIn
(SN\,1998S, Fassia et al. 2000;
%SN\,2005cf; Wang et al. 1999;
%SN\,1994i; Richmond et al. 1996;
SN\,2005gl, Gal-Yam et al. 2007).
\label{fig:LC_PTF09uj_all}}
\end{figure}

The {\it GALEX} photometry was carried out by performing aperture photometry
with a $10''$ radius around the SN host galaxy,
and subtracting its light as measured
in the reference image.
The reference image was constructed by combining the four
{\it GALEX} images of this field taken prior to the SN explosion,
between 2009 May 4 and May 14.
%and have a total integration time of 5442\,s.
% GALEX ref images
% Date      ExpTime
%04-05-2009  1428
%10-05       1271
%12-05       1331
%14-05       1412
%
Since GALEX uses photon-counting detectors
(i.e., individual photons are time tagged; Martin et al. 2005),
we had the opportunity
to look for flux variations on relatively short timescales.
In particular, we examined the earliest image
in which the SN was detected, which was taken on June 22 and had an
exposure time of 1364 seconds.
We extracted the time-tags of the 4597 photons found within $10''$ of the
SN and binned
these photons on timescales from 3 to 1000s.
We found no significant variations in flux as a function of time.
We note that about $40\%$ of these photons
originate from the SN and the rest are due to the host galaxy.
The {\em GALEX} {\it NUV} and P48 $R$-band light curves of PTF\,09uj
are presented in Figure~\ref{fig:LC_PTF09uj_all}.

%
%\begin{deluxetable*}{cccccccccccccc}
\begin{deluxetable}{lllll}
%\begin{deluxetable}{cccccccccccccc}
\tablecolumns{5}
\tablewidth{0pt}
\tablecaption{Observations of PTF\,09uj}
\tablehead{
\colhead{Telescope} &
\colhead{UTC 2009} &
\colhead{band} &
\colhead{Magnitude} &
%\colhead{Error} &
\colhead{$f_{\nu}$} \\
\colhead{}        &
\colhead{}        &
\colhead{}        &
\colhead{[AB mag]}&
%\colhead{[mag]}   &
\colhead{[$\mu$Jy]}     
}
\startdata
% Telescope   UTC 2009     band  AB-Mag    Err   F_{nu} [erg\,s$^{-1}$\,\cm^-2\,Hz^-1$]
P48         & Jun 02.3\tablenotemark{a}   & $R$\tablenotemark{b}    & $>20.5$  & $<23$ \\
            & Jun 23.30  & $R$    & $19.22\pm 0.13$  & $75$ \\
            & Jul 03.22  & $R$    & $19.27\pm 0.20$  & $71$ \\
            & Jul 07.20  & $R$    & $19.72\pm 0.25$  & $47$ \\
%            & Jul 09.23  & $R$    & 19.82  & 0.25  & $43$ \\
            & Jul 10.21  & $R$    & $20.08\pm 0.24$  & $34$ \\
            & Jul 12.24  & $R$    & $20.21\pm 0.32$  & $30$ \\
            & Jul 14.23  & $R$    & $20.37\pm 0.42$  & $26$ \\
            & Jul 16.26  & $R$    & $20.66\pm 0.31$  & $20$ \\
            & Jul 19.21  & $R$    & $20.89\pm 0.74$  & $16$ \\
            & Jul 22.20  & $R$    & $21.06\pm 0.30$  & $14$ \\
\hline
P60         & Jun 26.31  & $g$    & $18.17\pm0.07$   & $197$ \\
            & Jun 26.31  & $r$    & $18.31\pm0.04$   & $173$ \\
            & Jun 26.31  & $i$    & $18.42\pm0.05$   & $159$ \\
            & Jun 30.38  & $g$    & $18.69\pm0.07$   & $122$ \\
            & Jun 30.38  & $r$    & $18.71\pm0.05$   & $119$ \\
            & Jun 30.38  & $i$    & $18.70\pm0.11$   & $122$ \\
\hline
{\it GALEX} & Jun 20.36  & $NUV$  & $>21.7$          & $<7.5$\\
            & Jun 22.35  & $NUV$  & $19.32\pm 0.04$  & $67$\\
            & Jun 25.98  & $NUV$  & $17.80\pm 0.02$  & $274$\\
\hline
Lick\tablenotemark{c}&Jun 28.27& spec&             &   \\
%            &            & $g$  & $18.2\pm 0.3$    & $192$ \\
            &            & $r$  & $18.4\pm 0.3$    & $159$ 
%            &            & $i$  & $18.7\pm 0.3$    & $122$ \\
%     F_NUV=get_filter('GALEX','NUV');
%     [Const,SmJy,F_lambda,F_nu]=mag_jy_conv(F_NUV.nT{1},F_NUV.eff_wl{1},
% 0,'AB',[21.8;19.44;17.92]); [F_lambda, F_nu].*4.*pi.*(Dist.*3.08e18).^2
\enddata
\tablenotetext{a}{The last P48 non-detection before the discovery.}
\tablenotetext{b}{All of the P48 observations were conducted using the
  Mould $R$-band filter. Photometry was measured in the combined images of
  the same field taken each night (usually two).}
\tablenotetext{c}{The magnitude from Lick observatory is based on
  synthetic photometry of the spectrum using the code described by
  Poznanski et al. (2002).}
\tablecomments{$f_{\nu}$ is calculated at 2316\,\AA, 4718\,\AA,
  6184\,\AA, and 7499\,\AA\ for the $NUV$, $g$, PTF $R$/$r$, and
  $i$ bands, respectively. Magnitude uncertainties include (in
  quadrature) absolute calibration errors of 0.099\,mag for the PTF
  $R$-band measurements, and 0.071, 0.037, and 0.029\,mag for the P60 $g$,
  $r$, and $i$ bands, respectively.
  An aperture correction of 0.12\,mag was applied to the
  GALEX-$NUV$ magnitudes (Morrissey et al. 2007).}
\label{tab:Obs}
\end{deluxetable}
%\end{deluxetable}

%

\subsection{Spectroscopy}

We observed PTF\,09uj
with an exposure time of 1800\,s with the Kast double
spectrograph (Miller \& Stone 1993) mounted at the Cassegrain focus of
the Shane 3-m telescope at Lick
Observatory; the 5500\,\AA\ dichroic was employed.
On the blue arm, we used the 600\,lines\,mm$^{-1}$ grism blazed at 4310\,\AA~to
provide spectral coverage of 3500--5550\,\AA\ and a dispersion of
1.02\,\AA\,pixel$^{-1}$, while on the red arm we used the
300\,lines\,mm$^{-1}$ grating blazed at 7500\,\AA\ for a wavelength range of
5400--10000\,\AA\ and a dispersion of 4.60\,\AA\,pixel$^{-1}$.

\begin{figure}
\centerline{\includegraphics[width=8.5cm]{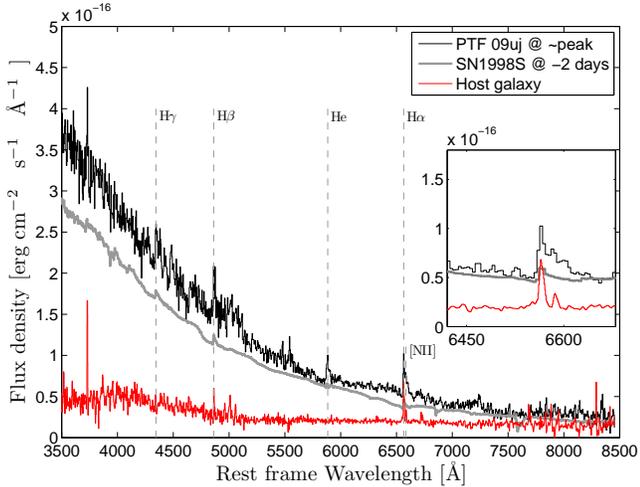}}
\caption{The spectrum of PTF\,09uj ({\it black line}; upper spectrum)
and its host galaxy ({\it red line}; lower spectrum).
For comparison, the scaled spectrum of the Type IIn SN\,1998S
(Fassia et al. 2001)
two days prior to maximum light is shown just
below the spectrum of PTF\,09uj ({\it gray line}).
The inset shows the H$\alpha$ line in detail.
\label{fig:Spec_PTF09uj}}
\end{figure}

The Lick spectrum was reduced using standard routines in
IRAF\footnote{IRAF is distributed by the National Optical Astronomy
Observatory, which is operated by the Association for Research in
Astronomy, Inc., under cooperative agreement with the National Science
Foundation.} (details provided by Cenko et al. 2008).
%Telluric absorption features were removed using the
%%well-exposed
%continuum of spectrophotometric
%standard stars (e.g., Wade \& Horne 1988; Matheson et al. 2000).
Flux calibration was performed relative
to the standard stars BD\,+28$^{\circ}$\,4211 (blue side) and 
BD\,+26$^{\circ}$\,2606 (red side).
The Lick spectrum of PTF\,09uj is shown in Figure~\ref{fig:Spec_PTF09uj}.
Given the lack of \ion{Na}{1} absorption lines in
the SN spectrum and the low Galactic extinction
toward this SN ($E_{B-V} = 0.011$ mag; Schlegel et al. 1998),
we do not correct for extinction.

%

%H\alpha FWHM 10.2\,\AA flux = 1.4e-15  EW=-9.7
%H\beta  FWHM  5.09            5.7e-16  EW=-1.9
%
%H\alpha/ H\beta= 2.5
%
%
%host:
%H\alpha FWHM  3.2\,\AA flux = 8.7e-16  EW=-15.5
%H\beta  FWHM  3.5             2.5e-16  EW=-4.0
%
%H\alpha/ H\beta= 3.5

\section{Interpretation}
\label{sec:Int}

The spectrum of PTF\,09uj, which was taken around peak light, shows 
a blue continuum, with narrow H$\alpha$ emission and no
prominent, broad absorption features.
This ``narrow'' line is actually broader than the H$\alpha$ line from
the host galaxy and shows a hint of
a P-Cygni profile (Fig.~\ref{fig:Spec_PTF09uj} inset).
The spectrum also exhibits a narrow He\,I emission line
(measured rest wavelength 5884\,\AA, corresponding to He\,I $\lambda$5876)
which is not present in the host-galaxy spectrum.
These observations
suggest that PTF\,09uj is a SN~IIn 
enshrouded with a dense CSM.

However, the $e$-folding decline rate of the SN flux
is about 10\,days.
This is faster than the steepest declining SNe~IIn 
previously known, such as SN\,1998S (Fassia et al. 2000),
SN\,2005gl (Gal-Yam et al. 2007), and SN\,2005ip (e.g., Smith et al. 2009);
compare with the light curves of the first 
two in Figure~\ref{fig:LC_PTF09uj_all}.

Another possible difference between PTF\,09uj and SN\,1998S is the spectra.
While the SNe spectra shown in Figure~\ref{fig:Spec_PTF09uj} were both taken around maximum light,
PTF\,09uj evolved faster and therefore these spectra probably do not correspond to
the same epoch after explosion. In order to compare the spectra of the SNe
taken at the same epoch after explosion
it is probably more adequate to inspect the spectrum of SN\,1998S
taken $13$ days prior to maximum light (Fassia et al. 2001).
This earlier spectrum of SN\,1998S
is different, with broader emission lines and strong ``Wolf-Rayet''-like features
(e.g., \ion{C}{3}, \ion{N}{3}).

\subsection{Shock Breakout in a Stellar Wind}
\label{SB}

The fast rise in UV light 
and the high peak luminosity 
($\nu f_{\nu} \approx 3\times10^{43}$\,erg\,s$^{-1}$)
motivates us to consider a model of 
a shock breakout which takes place within
a dense, optically thick, stellar wind
(see also Falk \& Arnett 1977; Waxman et al. 2007).
The blue continuum in the visible-light spectrum of PTF\,09uj
suggests that the emission is optically thick.
Fitting a black-body curve to the $NUV$, $g$, $r$, and 
$i$-band photometry, obtained on June 25--26, we find a 
best-fit temperature of $\approx 1.7 \times 10^{4}$\,K
(with r.m.s. of 0.13\,magnitudes).
We note that if  line blanketing is effecting the spectrum,
the true effective temperature could be even higher.

\begin{figure}
\centerline{\includegraphics[width=8.5cm]{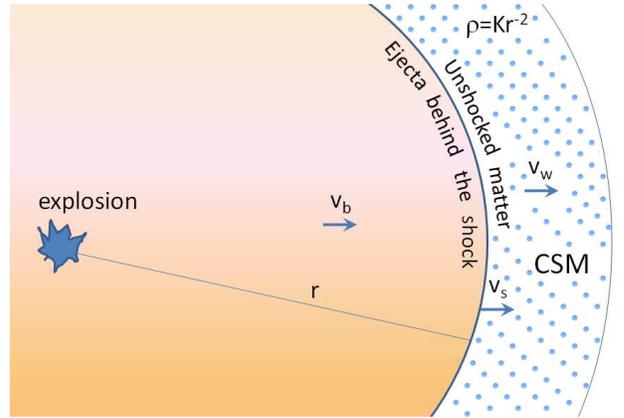}}
\caption{A sketch of our model for PTF\,09uj.
Fast ejecta from the SN explosion are interacting with a CSM
having a wind-like density profile.
Here $v_{{\rm w}}$ is the wind speed.
See definitions of variables
in \S\ref{SB}. 
\label{fig:PTF09uj_scetch}}
\end{figure}

In the framework of the model considered here
(see sketch in Fig.~\ref{fig:PTF09uj_scetch}),
the rising UV emission is
due to a shock breakout within an optically thick wind.
Some or most of the visible-light emission
at later times is caused by
diffusion of the shock-deposited energy.

We now calculate the properties of the shock and ejecta that are needed
in order to explain the observations.
In this calculation we use the observed peak luminosity and the rise time of the
SN to calculate various parameters
(i.e., mass, velocity, temperature).

The thickness of a radiation-mediated shock, $\tau_{{\rm s}}$,
in units of the Thomson optical depth (e.g., Weaver 1976) is given by
\begin{equation}
\tau_{{\rm s}}\approx c/v_{{\rm s}}.
%\approx \kappa r \rho.
\label{tau_s}
\end{equation}
Here $c$ is the speed of light and $v_{{\rm s}}$ is the
upstream ejecta (and shock) speed.
A radiation-mediated shock ``breaks down'' or ``breaks out''
(i.e., radiation escapes ahead of the shock)
when photons diffuse ahead of the shock faster than
the shock propagates.
For a wind-density profile
$\rho(r) = Kr^{-2}$ where $K$ is a normalization constant and $r$
is distance from the center,
the photon diffusion time from $r$ to $2r$ is
$t_{{\rm diff}}=\kappa \rho r^{2}/c=\kappa K / c$,
independent of $r$,
while the shock propagation time is $r/v_{{\rm s}}$,
growing with $r$.
Thus, the shock breaks down when it reaches $r_{{\rm break}}=\kappa K v_{{\rm s}}/c$.
At this point, photons would diffuse and escape the wind on
a time scale equal to $t_{{\rm diff}}$ multiplied by a log correction
factor $\ln(c/v_{{\rm s}})$.

An alternative definition of the breakout radius is the radius at
which the thickness of the shock is comparable to the scale
height of the density variation. Since for a wind profile
this scale is $r$, with a corresponding optical depth
$\tau=\kappa \rho r=\kappa K/r$, this happens
at $\tau_{{\rm s}}=\kappa K/r=c/v_{{\rm s}}$
or $r_{{\rm break}}=\kappa K v_{{\rm s}}/c$,
the same as obtained above.
For small radii $r$,
the shock expansion time scale ($t_{\rm exp} = r/v$)
is shorter than $t_{{\rm diff}}$,
and breakout takes place when $t_{{\rm diff}} \approx t_{\rm exp}$.
We further note that $v_{{\rm s}}$ corresponds to the velocity 
of the faster parts of the shock.
In \S\ref{BulkVel} we will derive the ratio of this velocity to the 
bulk velocity of the ejecta ($v_{{\rm b}}$).

Assuming a radiation-dominated shock,
the downstream temperature of the shock can be estimated
by comparing the radiation energy density (assuming black-body 
radiation) with the kinetic energy per unit volume:
\begin{equation}
a T^{4} \approx \frac{7}{2}\rho_{{\rm s}} v_{{\rm s}}^{2} \approx \frac{7c}{2t\kappa},
\label{Tshock}
\end{equation}
where $a$ is the radiation constant,
%($=4\sigma_{B}/c$),
%=7.5657\times10^{-15}$\,erg\,cm$^{-3}$\,K$^{-4}$)
$t$ is the time since the explosion,
$\kappa$ is the opacity,
and $\rho_{{\rm s}}$ is the density of the stellar wind at the shock breakout.
Note that the right-hand side of Equation~\ref{Tshock} is derived
assuming a wind-density profile $\rho(r) \propto r^{-2}$
along with Equation~\ref{tau_s}
and $\tau=\int_{r_{{\rm s}}}^{\infty}{\kappa \rho(r) dr}=\kappa \rho r_{{\rm s}}$.
The factor of $7$ arises from the
shock's compression ratio $(\gamma+1)/(\gamma-1)=7$,
with an adiabatic index of $\gamma=4/3$.
Here $r_{{\rm s}}$ is the radius at which the shock breakout takes place.

In this framework, the
kinetic energy of the explosion is converted into radiation
at a rate of
$L\approx 2\pi r_{{\rm s}}^{2} v_{{\rm s}} \rho_{{\rm s}} v_{{\rm s}}^{2}$.
%where $r_{{\rm s}}$ is the distance from the star to the shell at which the breakout
%occurs\footnote{This is correct because at the time of breakout, the photon diffusion time is the same as the shock propagation time.}.
%
Here we assume that
the shock is optically thick and that the
emission spectrum is roughly
represented by a black-body spectrum.
The assumption of black-body emission is justified, since for the shock
velocity we infer, $v/c\approx0.03$ (see Eq.~\ref{v} below),
there is no departure
from thermal equilibrium. Such departure is expected only for higher
velocity shocks, $v/c\gtorder 0.1$ (Weaver 1976; Katz et al. 2010).

Based on Equation~\ref{tau_s}, and assuming
we are observing at wavelength $\lambda$ in
the Rayleigh-Jeans tail,
\begin{equation}
L_{\lambda}\approx \frac{v_{{\rm s}}}{c}L_{\lambda}^{BB} 
\approx 4 \pi^{2} r_{{\rm s}}^{2} v_{{\rm s}} \frac{2k_{B}T}{\lambda^{4}},
\label{l_l}
\end{equation}
where $L_{\lambda}^{BB}$ is the black-body total specific luminosity
and $k_{B}$ is the Boltzmann constant.
%The right hand side of Equation~\ref{l_l} assumes
%that the observations are conducted
%well within the Rayleigh-Jeans tail.

By solving Equations \ref{Tshock} and \ref{l_l},
along with $r=v_{{\rm s}}t$,
we find that
\begin{equation}
v_{{\rm s}}  \approx 1.3\times10^{4}
              L_{\lambda, {\rm 6e38}}^{1/3}
              \lambda_{6200}^{4/3}
              \kappa_{0.34}^{1/12}
              t_{7}^{-7/12}
              \,{\rm km\,s}^{-1},
\label{v}
\end{equation}
where $L_{\lambda, {\rm 6e38}}$ is the specific luminosity in units
of $6 \times 10^{38}$\,erg\,s$^{-1}$\,\AA$^{-1}$
(equivalent to the luminosity measured on 2009 June 23, close 
to the first {\em GALEX} detection), $\lambda_{6200}$ is the 
wavelength at which the luminosity is measured in units of 6200\,\AA,
$\kappa_{0.34}$ is the opacity in units of 0.34\,cm$^{2}$\,g$^{-1}$ 
(assuming the opacity of completely ionized solar composition),
and $t_{7}$ is the time between the explosion and the measurement of
$L_{\lambda}$, in units of $7$\,days.
The distance from the star to the shell in which the breakout occurs is
\begin{equation}
r_{{\rm s}} = v_{{\rm s}}t \approx 8.2\times10^{14} 
              L_{\lambda, {\rm 6e38}}^{1/3}
              \lambda_{6200}^{4/3}
              \kappa_{0.34}^{1/12}
              t_{7}^{5/12}
              \,{\rm cm}.
\label{r}
\end{equation}
This radius is larger than the size of a typical red supergiant, indicating
that our assumption that the shock occurs ``outside'' the star,
in a wind-density profile, is justified.
The density is given by
\begin{equation}
n = \frac{\rho_{{\rm s}}}{m_{{\rm p}}} \approx 4.8\times10^{10}
              L_{\lambda, {\rm 6e38}}^{-2/3}
              \lambda_{6200}^{-8/3}
              \kappa_{0.34}^{-7/6}
              t_{7}^{1/6}
              \,{\rm cm}^{-3},
\label{rho}
\end{equation}
where $m_{{\rm p}}$ is the proton mass,
and we also have
\begin{equation}
T    \approx 9.1\times10^{4}
              \kappa_{0.34}^{-1/4}
              t_{7}^{-1/4}
              \,{\rm K}.
\label{T}
\end{equation}
Here, $T$ is the effective temperature at the time
in which we measured $L_{\lambda}$.

In addition, we can estimate the rise time of the SN light curve.
In the case that the scale height of the material lying ahead of
the shock is negligible compared with the radius at which the shock occurs
($r_{{\rm s}}$; i.e., shock takes place at the edge of the star),
the rise time is $\sim r_{{\rm s}}/c$
(up to a log correction factor mentioned earlier).
However, in the scenario discussed here
there is a significant amount of material ahead of the shock and
the rise time is $t_{{\rm rise}}\approx r_{{\rm s}}/v_{{\rm s}} = 7t_{7}$\,day.
This rise time is therefore slower than the rise time
seen in some of the known shock breakout events (e.g., Gezari et al. 2008).
In this scenario, after maximum light, the light curve
is expected to decay exponentially on the diffusion
time scale ($t_{{\rm diff}}$; Falk \& Arnett 1977),
which in our case is about a week.
This is similar to the observed decay rate.

The total mass in the fast ejecta within radius $r_{{\rm s}}$ 
is $\int_{r_{*}}^{r_{{\rm s}}}{4\pi r^{2}\rho(r)dr}$.
Here $r_{*}$ is the star radius.
Assuming the wind density is $\rho(r)\propto r^{-2}$ and 
$r/r_{*}\gg 1$, we find
\begin{equation}
M \approx 4\pi \rho r_{{\rm s}}^{3} \approx 0.3
              L_{\lambda, {\rm 6e38}}^{1/3}
              \lambda_{6200}^{4/3}
              \kappa_{0.34}^{-11/12}
              t_{7}^{17/12}
              \,{\rm M}_{\odot}.
\label{M}
\end{equation}
Note that the total mass in the ejecta may be larger
(see \S\ref{BulkVel}), and therefore this equation
provides only a lower limit to the mass.
Next, the mass loss rate from the progenitor prior to the explosion is
\begin{equation}
\dot{M} \approx \frac{M v_{w,100}}{r} \approx  0.1
              \kappa_{0.34}^{-1}
              t_{7}
              v_{w,100}^{-1}
              \,{\rm M}_{\odot}\,{\rm yr}^{-1},
\label{Mdot}
\end{equation}
where $v_{w,100}$ is the progenitor wind velocity in 
units of 100\,km\,s$^{-1}$.
This mass-loss rate is required to persist for 
$r_{{\rm s}}/v_{w} \approx 10$\,yr prior
to the explosion.

Another important property is the kinetic energy 
in the faster parts of the shock,
%(i.e., not only the illuminated energy),
which is roughly given by
\begin{eqnarray}
E   & \approx & 4\pi r_{{\rm s}}^{2}\frac{r_{{\rm s}}}{7} aT^{4} \approx  
\frac{4}{2}\pi r_{{\rm s}}^{3} \rho v_{{\rm s}}^{2} \cr
    & \approx &
           5\times10^{50}
              L_{\lambda, {\rm 6e38}}
              \lambda_{6200}^{4}
              \kappa_{0.34}^{-3/4}
              t_{7}^{1/4}
              \,{\rm erg}.
\label{E}
\end{eqnarray}
As before the factor of $7$ arises from the shock's compression ratio.

We note that if we relax the Rayleigh-Jeans approximation
in Equation~\ref{l_l}, and solve these equations using the full
Planck formula,
the solution does~not change
considerably\footnote{The values of $v_{{\rm s}}$, $r_{{\rm s}}$, $t_{{\rm rise}}$, and 
$M$ are changed by a factor of $1.04$; $n$ is changed by a factor 
of $0.92$; $E$ by $1.14$; and $T$ and $\dot{M}$ remain the same.}.
Therefore, for simplicity,
we choose to show here the approximate solution.

Based on the temperature that we derive in Equation~\ref{T}, the expected
{\it NUV} luminosity is
$2.1\times10^{40}$\,erg\,s$^{-1}$\,\AA$^{-1}$
(given by the Planck function multiplied by $v_{{\rm s}}/c$).
This is about $1.5$ times larger than observed.
However, the UV emission may be affected by
metal line blanketing,
and is therefore less reliable than the visible-light luminosity.

Furthermore, we note that our assumption of a radiation-mediated 
shock is justified
since in the case studied here the ratio of radiation energy density
to plasma thermal energy density is very large,
$aT^{4}/(nk_{B}T) \approx 10^{6}$. 

To summarize, based on our crude model we use the visible-light 
luminosity to derive the shock properties 
(e.g., $r_{{\rm s}}$, $v_{{\rm s}}$, $T$, $E$, $\dot{M}$).
The calculated $t_{{\rm rise}}$ is roughly consistent with the
observed rise time of the UV light curve.
Moreover, this model naturally explains the high observed luminosity in the
{\it NUV} band.

\subsection{The Bulk Velocity}
\label{BulkVel}
The velocity $v_{{\rm s}}$ we used so far corresponds to the faster parts
of the ejecta.
However, this does~not necessarily represent
most of the energy in the ejecta.
As a sanity check, here we estimate the 
bulk velocity of the ejecta.

The shock that accelerates the ejecta gives more energy to slower shells.
The energy as a function of velocity is given by
\begin{equation}
E(v)=E_{{\rm b}} \Big( \frac{v}{v_{{\rm b}}} \Big)^{-x},
\label{Ev}
\end{equation}
where $E_{{\rm b}}$ is the bulk kinetic energy of the ejecta,
$v_{{\rm b}}$ is the bulk velocity,
and $x=5(1+3n/5)/n$, where $n=3/2$ ($3$), $x=19/3$ ($14/3$),
for convective (radiative)
envelopes\footnote{In the notation of Matzner \& McKee (1999), 
$x = (n+1)/(n\beta_{1})-2$, where $\beta_{1}=1/5$.}
(Matzner \& McKee 1999).
A shell of velocity $v$ is decelerated when the energy in the
shocked wind, $E_{w}$, equals the energy in the decelerated shell, $E(v)$.
Here $E_{w}=4\pi K r v^{2}$ and $E(v)=M(v)v^{2}$.
Note that this includes the internal energy in the shocked wind
(assuming the internal energy roughly equals the kinetic energy).
This gives the deceleration radius
\begin{equation}
r(v)=\frac{E_{\rm b}}{4\pi K v_{{\rm b}}^{2}} \Big( \frac{v}{v_{{\rm b}}} \Big)^{-2-x}.
\label{rv}
\end{equation}
The wind optical depth at this radius is $\kappa \rho r$, or
\begin{equation}
\tau(v)=\frac{4\pi \kappa K^{2} v_{{\rm b}}^{2}}{E_{{\rm b}}} \Big(\frac{v}{v_{0}}\Big)^{2+x}.
\label{tauv}
\end{equation}
As long as $\tau\gtorder c/v_{{\rm s}}$, the fast shell is decelerated and
overtaken by the slower, more energetic, shells behind it.
Breakout occurs when $\tau(v) \approx c/v_{{\rm s}}$, 
which together with Equation~\ref{tauv}
and assuming $E_{{\rm b}}=M_{{\rm b}} v_{{\rm b}}^{2}/2$ gives
%\begin{equation}
%(v/v_0)=[(c/v)(E_0/(4\pi*kappa*A^2*v_0^2)]^{1/(2+x)}.
%\label{vv0}
%\end{equation}
%Using E_0=M_0 v_0^2/2, this gives
\begin{equation}
\frac{v_{{\rm s}}}{v_{{\rm b}}} = \Big( \frac{c}{v_{{\rm s}}} 
\frac{M_{{\rm b}}}{8\pi \kappa K^{2}} \Big)^{1/(2+x)},
\label{vv0}
\end{equation}
%   (v/v_0)=[(c/v)(M_0/(8\pi*kappa*A^2)]^{1/(2+x)}.
where $M_{{\rm b}}$ is the total mass in the ejecta.
Since $1/(x+2) \ll 1$, the inferred value of $v_{{\rm s}}/v_{{\rm b}}$
is insensitive to the exact values of $v_{{\rm s}}$ and $K$, and to
the unknown value of $M_{\rm b}$.
Assuming $M_{{\rm b}}=1$\,M$_{\odot}$, $v_{{\rm s}}=10^{4}$\,km\,s$^{-1}$,
and $K=\dot{M}/ (4\pi v_{w}) \approx5\times10^{16}$\,g\,cm$^{-1}$
(corresponding to $\dot{M}=0.1$\,M$_{\odot}$\,yr$^{-1}$
and $v_{w}=100$\,km\,s$^{-1}$), 
we get $v_{{\rm s}}/v_{{\rm b}}\approx 1.1$ ($1.2$)  for convective 
(radiative) envelopes.

This analysis suggests that Equations~\ref{v}, \ref{M}, and \ref{E}
are reasonable approximations of the bulk properties of the ejecta.
Specifically, if the total mass of the ejecta is an order
of magnitude larger than that given by Equation~\ref{M},
then the total kinetic energy of the SN
will exceed $5\times10^{51}$\,erg,
which is unlikely (at least for the garden variety of SNe).

\section{Discussion}
\label{sec:Disc}

We present the discovery of PTF\,09uj, which was serendipitously
observed by {\it GALEX}  at early times after the explosion.
The spectrum of the SN and the bright UV signal
suggest that this was a SN~IIn powered by the
diffusion of the shock energy and
interaction of the ejecta with a dense CSM 
($n \approx 10^{10}$\,cm$^{-3}$).
This interpretation is consistent with both the fast rise
of the UV light curve and the UV luminosity.
The observed fast rise cannot be easily explained
unless the progenitor is embedded in an optically thick wind.
Moreover, a shock breakout from a stellar photosphere
cannot generate such a bright UV signal
(e.g., Rabinak \& Waxman 2010).

Based on simple modeling, we suggest that
prior to the explosion the progenitor went through a phase of
high mass-loss rate, with $\dot{M} \approx 0.1$\,M$_{\odot}$\,yr$^{-1}$.
The radius of the radiating region and the fast decay
of the SN are suggestive of an episodic high mass-loss rate
with a duration of about several years prior to the explosion.
Our model suggests
that the total mass of the ejecta is relatively low, roughly
1\,M$_{\odot}$.
We stress that this is an order-of-magnitude estimate; the 
total ejected mass could be as high as a 
few solar masses, but 
probably not on the high end ($\gtorder 10$\,M$_{\odot}$)
of the ejected mass expected in typical SNe~II.

The low ejecta mass, if true, may be due to one of the following:
(i) most of the mass of the progenitor was not expelled by the SN
explosion and it is in a form of a compact remnant;
or (ii) most of the mass of the progenitor was expelled (e.g., wind)
prior to the explosion.
This is in contrast to more energetic SN explosions
whose luminosity is powered by interaction with a dense CSM
(e.g., Benetti et al. 2006; Ofek et al. 2007; Smith et al. 2007)

PTF is a wide and shallow
survey in which about 3000\,deg$^{2}$ are actively surveyed
at a given time down to a limiting magnitude of $\sim 21$.
{\em GALEX} sensitivity allows high signal-to-noise 
ratio detections of SN shock-breakout flashes at a redshift range
which is similar to that probed by PTF.
Specifically, we estimate that PTF should find several SNe each year
for which {\it GALEX} early observations will be available.
This estimate is based on the number of 
PTF SN discoveries, during 2009, which had
{\it GALEX} serendipitous observations
between 0 to 30 days prior to the PTF discovery.
%
%We note that additional results are expected from 
% Pan-STARRS1 (Hodapp et al. 2004).
%Although given the narrow and deep nature of this survey, 
% it is optimized toward
%higher redshift events with luminous UV signals
%(e.g., Botticella et al. 2010).
%

\acknowledgments

We thank an anonymous referee for useful comments.
E.O.O. and D.P. are supported by an Einstein fellowship.
S.B.C. and A.V.F. acknowledge generous financial assistance
from Gary \& Cynthia Bengier, the Richard \& Rhoda Goldman Fund, 
NASA/{\it Swift} grants NNX09AL08G and NNX10AI21G, and NSF grant AST-0908886.
A.G. acknowledges support by the Israeli and the US-Israel Binational Science 
Foundations, an EU/IRG fellowship, the Benoziyo Center for Astrophysics,
and the Peter and Patricia Gruber Awards.
The National Energy Research Scientific Computing Center, which is supported by
the Office of Science of the U.S. Department of Energy under Contract No.
DE-AC02-05CH11231, provided staff, computational resources, and data storage for
this project. P.E.N. acknowledges support from the US Department of Energy Scientific
Discovery through Advanced Computing program under contract DE-FG02-06ER06-04.
J.S.B.'s work on PTF was supported by NSF/OIA award AST-0941742
(``Real-Time Classification of Massive Time-Series Data Streams").
L.B. and K.S. are supported by the NSF under
grants PHY 05-51164 and AST 07-07633.

\end{document}